\documentclass[
    ,final            
  ]
  {aipproc}
\usepackage{epsfig}
\layoutstyle{6x9}

\def\eg{{e.g., }}
\def\ie{{i.e., }}
\def\etal{{et al., }}

\def\etc{{etc.}}

\def\'{^{\prime}}

\def\avrg#1{{\langle #1 \rangle}}

\def\hmpc{{\, {\rm h}^{-1}~\rm Mpc}}

\def\kms{{\rm~km~s^{-1}}}

\def\kpc{{\rm~kpc}}
\def\mpc{{\rm~Mpc}}

\def\spose#1{\hbox to 0pt{#1\hss}}
\def\lta{\mathrel{\spose{\lower 3pt\hbox{$\mathchar"218$}}
     \raise 2.0pt\hbox{$\mathchar"13C$}}}
\def\gta{\mathrel{\spose{\lower 3pt\hbox{$\mathchar"218$}}
     \raise 2.0pt\hbox{$\mathchar"13E$}}}
\def\ge{\mathrel{\spose{\lower 3pt\hbox{$-$}}
     \raise 2.0pt\hbox{$\mathchar"13E$}}}
\def\le{\mathrel{\spose{\lower 3pt\hbox{$-$}}
     \raise 2.0pt\hbox{$\mathchar"13C$}}}

\newcommand{\cbi}{CBI}

\begin{document}

\title{The Cosmic Microwave Background \& Inflation, \\
 Then \& Now}

\author{J. Richard Bond$^{1}$, Carlo Contaldi$^{1}$, Dmitry
Pogosyan$^{2}$, Brian Mason$^{3,4}$, Steve Myers$^{4}$, Tim
Pearson$^{3}$, Ue-Li Pen$^{1}$, Simon Prunet$^{5,1}$, Tony
Readhead$^{3}$, Jonathan Sievers$^{3}$} {address={ 1. CIAR Cosmology
Program, Canadian Institute for Theoretical Astrophysics, \\ Toronto,
Ontario, Canada \\ 2. Physics Department, University of Alberta,
Edmonton, Alberta, Canada \\ 3. Astronomy Department, California
Institute of Technology, Pasadena, California, USA \\ 4. National
Radio Astronomy Observatory, Socorro, New Mexico, USA \\ 5. Institut
d'Astrophysique de Paris, Paris, France \\
        }}

\begin{abstract}
The most recent results from the Boomerang, Maxima, DASI, CBI and VSA
CMB experiments significantly increase the case for accelerated
expansion in the early universe (the inflationary paradigm) and at the
current epoch (dark energy dominance). This is especially so when
combined with data on high redshift supernovae (SN1) and large scale
structure (LSS), encoding information from local cluster abundances,
galaxy clustering, and gravitational lensing. There are ``7 pillars of
Inflation'' that can be shown with the CMB probe, and at least 5, and
possibly 6, of these have already been demonstrated in the CMB data:
(1) the effects of a large scale gravitational potential, demonstrated
with COBE/DMR in 1992-96; (2) acoustic peaks/dips in the angular power
spectrum of the radiation, which tell about the geometry of the
Universe, with the large first peak convincingly shown with Boomerang
and Maxima data in 2000, a multiple peak/dip pattern shown in data
from Boomerang and DASI (2nd, 3rd peaks, first and 2nd dips in 2001)
and from CBI (2nd, 3rd, 4th, 5th peaks, 3rd, 4th dips at 1-sigma in
2002); (3) damping due to shear viscosity and the width of the region
over which hydrogen recombination occurred when the universe was
400000 years old (CBI 2002); (4) the primary anisotropies should have
a Gaussian distribution (be maximally random) in almost all
inflationary models, the best data on this coming from Boomerang; (5)
secondary anisotropies associated with nonlinear phenomena subsequent
to 400000 years, which must be there and may have been detected by CBI
and another experiment, BIMA.  Showing the 5 ``pillars'' involves
detailed confrontation of the experimental data with theory; \eg (5)
compares the CBI data with predictions from two of the largest
cosmological hydrodynamics simulations ever done. DASI, Boomerang and
CBI in 2002, AMiBA in 2003, and many other experiments have the
sensitivity to demonstrate the next pillar, (6) polarization, which
must be there at the $\sim 7\%$ level. A broad-band DASI detection
consistent with inflation models was just reported. A 7th pillar,
anisotropies induced by gravity wave quantum noise, could be too small
to detect. A minimal inflation parameter set,
$\{\omega_b,\omega_{cdm},\Omega_{tot}, \Omega_Q,w_Q,n_s,\tau_C,
\sigma_8\}$, is used to illustrate the power of the current
data. After marginalizing over the other cosmic and experimental
variables, we find the current CMB+LSS+SN1 data give $\Omega_{tot} =
1.00^{+.07}_{-.03}$, consistent with (non-baroque) inflation
theory. Restricting to $\Omega_{tot}=1$, we find a nearly scale
invariant spectrum, $n_s =0.97^{+.08}_{-.05}$. The CDM density,
$\omega_{cdm}=\Omega_{cdm}{\rm h}^2 =.12^{+.01}_{-.01}$, and baryon
density, $\omega_b\equiv \Omega_b {\rm h}^2 = .022^{+.003}_{-.002}$,
are in the expected range. (The Big Bang nucleosynthesis estimate is
$0.019\pm 0.002$.)  Substantial dark (unclustered) energy is inferred,
$\Omega_Q \approx 0.68 \pm 0.05$, and CMB+LSS $\Omega_Q$ values are
compatible with the independent SN1 estimates. The dark energy
equation of state, crudely parameterized by a quintessence-field
pressure-to-density ratio $w_Q$, is not well determined by CMB+LSS
($w_Q < -0.4$ at 95\% CL), but when combined with SN1 the resulting
$w_Q < -0.7$ limit is quite consistent with the $w_Q$=$-1$
cosmological constant case.
\end{abstract}

\maketitle

\section{A Synopsis of CMB Experiments}

We are in the midst of a remarkable outpouring of results from the CMB
that has seen major announcements in each of the last three years,
with no sign of abatement in the pace as more experiments are
scheduled to release analyses of their results. This paper is an
update of \cite{capp00} to take into account how the new data have
improved the case for primordial acceleration, and for acceleration
occurring now. The simplest inflation models are strongly preferred by
the data. This does not mean inflation is proved, it just fits the
available information better than ever. It also does not mean that
competitor theories are ruled out, but they would have to look awfully
like inflation for them to work. As a result many competitors have now
fallen into extreme disfavour as the data have improved.

{\bf The CMB Spectrum:} The CMB is a nearly perfect blackbody of
$2.725 \pm 0.002\, K$ \cite{matherTcmb}, with a $3.372 \pm 0.007\, mK$
dipole associated with the 300 $\kms$ flow of the earth in the CMB,
and a rich pattern of higher multipole anisotropies at tens of $\mu$K
arising from fluctuations at photon decoupling and later. Spectral
distortions from the blackbody have been detected in the COBE FIRAS
and DIRBE data. These are associated with starbursting galaxies due to
stellar and accretion disk radiation downshifted into the infrared by
dust then redshifted into the submillimetre; they have energy about
twice all that in optical light, about a tenth of a percent of that in
the CMB. The spectrally well-defined Sunyaev-Zeldovich (SZ) distortion
associated with Compton-upscattering of CMB photons from hot gas has
not been observed in the spectrum. The FIRAS 95\% CL upper limit of $
6.0 \times 10^{-5}$ of the energy in the CMB is compatible with the
$\lta 10^{-5}$ expected from clusters, groups and filaments in
structure formation models, and places strong constraints on the
allowed amount of earlier energy injection, \eg ruling out mostly
hydrodynamic models of LSS. The SZ effect has been well observed at
high resolution with very high signal-to-noise along lines-of-sight
through dozens of clusters. The SZ effect in random fields may have
been observed with the CBI and BIMA, again at high resolution, although
multifrequency observations to differentiate the signal from the CMB
primary and radio source contributions will be needed to show this.

{\bf The Era of Upper Limits:} The story of the experimental quest for
spatial anisotropies in the CMB temperature is a heroic one. The
original 1965 Penzias and Wilson discovery paper quoted angular
anisotropies below $10\%$, but by the late sixties $10^{-3}$ limits
were reached, by Partridge and Wilkinson and by Conklin and
Bracewell. As calculations of baryon-dominated adiabatic and
isocurvature models improved in the 70s and early 80s, the theoretical
expectation was that the experimentalists just had to get to
$10^{-4}$, as they did, \eg Boynton and Partridge in 73.  The only
signal found was the dipole, hinted at by Conklin and Bracewell in 73,
but found definitively in Berkeley and Princeton balloon experiments
in the late 70s, along with upper limits on the quadrupole. Throughout
the 1980s, the upper limits kept coming down, punctuated by a few
experiments widely used by theorists to constrain models: the small
angle 84 Uson and Wilkinson and 87 OVRO limits, the large angle 81
Melchiorri limit, early (87) limits from the large angle Tenerife
experiment, the small angle RATAN-600 limits, the $7^\circ$-beam
Relict-1 satellite limit of 87, and Lubin and Meinhold's 89
half-degree South Pole limit, marking a first assault on the peak.

Primordial fluctuations from which structure would have grown in the
Universe can be one of two modes: adiabatic scalar perturbations
associated with gravitational curvature variations or isocurvature
scalar perturbations, with no initial curvature variation, but
variations in the relative amounts of matter of different types, \eg
in the number of photons per baryon, or per dark matter particle, \ie
in the entropy per particle.  Both modes could be present at once, and
in addition there may also be tensor perturbations associated
with primordial gravitational radiation which can leave an imprint on
the CMB. The statistical distribution of the primordial fluctuations
determines the statistics of the radiation pattern; \eg the
distribution could be generically non-Gaussian, needing an infinity of
N-point correlation functions to characterize it. The special case
when only the 3D 2-point correlation function is needed is that of a
Gaussian random field; if dependent upon spatial separations only, not
absolute positions or orientations, it is homogeneous and
isotropic. The radiation pattern is then a 2D Gaussian field fully
specified by its correlation function, whose spherical transform is
the angular power spectrum, ${\cal C}_\ell$, where $\ell$ is the
multipole number. If the 3D correlation does not depend upon
multiplication by a scale factor, it is scale invariant. This does not
translate into scale invariance in the 2D radiation correlation, whose
features reflect the physical transport processes of the radiation
through photon decoupling.

The upper limit experiments were in fact highly useful in ruling out
broad ranges of theoretical possibilities. In particular adiabatic
baryon-dominated models were ruled out. In the early 80s, universes
dominated by dark matter relics of the hot Big Bang lowered
theoretical predictions by about an order of magnitude over those of
the baryon-only models. In the 82 to mid-90s period, many groups
developed codes to solve the perturbed Boltzmann--Einstein equations
when such collisionless relic dark matter was present. Armed with
these pre-COBE computations, plus the LSS information of the time, a
number of otherwise interesting models fell victim to the data: scale
invariant isocurvature cold dark matter models in 86, large regions of
parameter space for isocurvature baryon models in 87, inflation models
with radically broken scale invariance leading to enhanced power on
large scales in 87-89, CDM models with a decaying ($\sim {\rm keV}$)
neutrino if its lifetime was too long ($\gta 10 {\rm yr}$) in 87 and
91. Also in this period there were some limited constraints on
"standard" CDM models, restricting $\Omega_{tot}$, $\Omega_B$, and the
amplitude parameter $\sigma_8$. ($\sigma_8^2$ is a bandpower for
density fluctuations on a scale associated with rare clusters of
galaxies, $8\hmpc$, where ${\rm h}=H_0/(100 \kms \mpc^{-1})$.)

{\bf DMR and Post-DMR Experiments to April 1999:} The familiar motley
pattern of anisotropies associated with $2 \le \ell \lta 20$
multipoles at the $30 \mu K$ level revealed by COBE at $7^\circ$
resolution was shortly followed by detections, and a few upper limits
(UL), at higher $\ell$ in 19 other ground-based (gb) or balloon-borne
(bb) experiments --- most with many fewer resolution elements than the
600 or so for COBE. Some predated in design and even in data delivery
the 1992 COBE announcement. We have the intermediate angle SP91 (gb),
the large angle FIRS (bb), both with strong hints of detection before
COBE, then, post-COBE, more Tenerife (gb), MAX (bb), MSAM (bb),
white-dish (gb, UL), argo (bb), SP94 (gb), SK93-95 (gb), Python (gb),
BAM (bb), CAT (gb), OVRO-22 (gb), SuZIE (gb, UL), QMAP (bb), VIPER
(gb) and Python V (gb). A list valid to April 1999 with associated
bandpowers is given in \cite{BJK2000}, and are referred here as 4.99
data. They showed evidence for a first peak \cite{BJK2000}, although
it was not well localized. A strong first peak, followed by a sequence
of smaller peaks diminished by damping in the ${\cal C}_\ell$ spectrum
was a long-standing prediction of adiabatic models. For restricted
parameter sets, good constraints were given on $n_s$, and on
$\Omega_{tot}$ and $\Omega_\Lambda$ when LSS was added~\cite{bj99}.

{\bf TOCO, BOOMERANG \& MAXIMA:} The picture dramatically improved
over the 3 years since April 1999. In summer 99, the ground-based TOCO
experiment in Chile \cite{toco98}, and in November 99 the North
American balloon test flight of Boomerang \cite{mauskopf99}, gave
results that greatly improved first-peak localization, pointing to
$\Omega_{tot}\sim 1$. Then in April 2000 dramatic results from the
first CMB long duration balloon (LDB) flight, Boomerang
\cite{debernardis00,lange00}, were announced, followed in May 2000 by
results from the night flight of Maxima \cite{Maxima00}. Boomerang's
best resolution was $10.7^\prime \pm 1.4^\prime$, about 40 times
better than that of COBE, with tens of thousands of resolution
elements. (The corresponding Gaussian beam filtering scale in
multipole space is $\ell_s \sim 800$.) Maxima had a similar resolution
but covered an order of magnitude less sky.  In April 2001, the
Boomerang analysis was improved and much more of the data were
included, delivering information on the spectrum up to $\ell \sim
1000$ \cite{Netterfield01,deBernardis01}. Maxima also increased its
$\ell$ range~\cite{Maxima01}.

\begin{figure}
\includegraphics[width=4.5in]{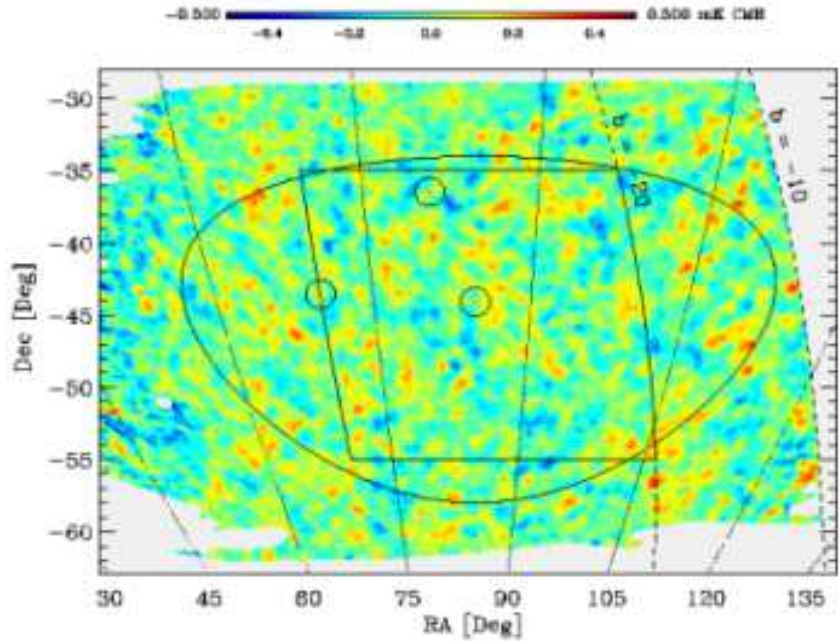} 
\vspace{5pt} \caption{The Boomerang 150 GHz bolometer map is shown in
the top figure. Of the entire 1800 square degrees covered, only the
interior 800 sq. degs. (within the ellipse) were used in
\cite{Netterfield01} and this analysis. In the April 2000 analysis
only one channel within the rectangle was used. In the Ruhl \etal 2002
analysis, essentially the entire region is covered. The three circles
show regions cut out of the analysis because they contain quasars with
emission at 150 GHz. The resolution is $10.7^\prime\pm 1.4^\prime$
{\it fwhm}. }
\label{fig:BOOMmap}
\end{figure}

Boomerang carried a 1.2m telescope with 16 bolometers cooled to 300 mK
in the focal plane aloft from McMurdo Bay in Antarctica in late
December 1998, circled the Pole for 10.6 days and landed just 50 km
from the launch site, only slightly damaged. Maps at 90, 150 and 220
GHz showed the same basic spatial features and the intensities were
shown to fall precisely on the CMB blackbody curve. The fourth
frequency channel at 400 GHz is dust-dominated. Fig.~\ref{fig:BOOMmap}
shows a 150 GHz map derived using four of the six bolometers at 150
GHz. There were 10 bolometers at the other frequencies.
Although Boomerang altogether probed 1800 square degrees,
the April 2000 analysis used only one channel and 440 sq. deg., and
the April 2001 analyses used 4 channels and the region in the ellipse
covering 800 sq. deg. That is the Boomerang data used in this
paper. In \cite{Ruhl02}, the coverage is extended to 1200 sq deg,
2.9\% of the sky.

Maxima covered a 124 square degree region of sky in the Northern
Hemisphere. Though Maxima was not an LDB, it did well because its
bolometers were cooled even more than Boomerang's, to 100 mK, leading
to higher sensitivity per unit observing time, it had a star camera so
the pointing was well determined, and, further, all frequency channels
were used in its analysis. 

{\bf DASI, CBI \& VSA:} DASI (the Degree Angular Scale
Interferometer), located at the South Pole, has 13 dishes of size
0.2m. Instead of bolometers, it uses HEMTs, operating at 10 frequency
channels spanning the band $26-36$GHz. An interferometer baseline
directly translates into a Fourier mode on the sky. The dish spacing
and operating frequency dictate the $\ell$ range. In DASI's case, the
range covered is $125 \lta \ell \lta 900$. 32 independent maps were
constructed, each of size $3.4^\circ$, the field-of-view (fov). The
total area covered was 288 sq. deg. DASI's spectacular results were
also announced in April 2001, unveiling a spectrum close to that
reported by Boomerang at the same time. The two results together
reinforced each other and lent considerable confidence to the emerging
${ \cal C}_\ell$ spectrum in the $\ell <1000$ regime.

CBI (the Cosmic Background Imager), based at 16000 feet on a high
plateau in Chile, is the sister experiment to DASI. It has 13 0.9m
dishes operating in the same HEMT channels as DASI. The instrument
measures 78 baselines simultaneously. The larger dishes by a factor
of 4 and longer baselines imply higher resolution by about the same
factor: the CBI results reported in May 2002 go to $\ell $ of 3500, a
huge increase over Boomerang, Maxima and DASI. Only the analyses of
data from the year 2000 observing campaign were reported. During 2000,
\cbi\ covered three deep fields of diameter roughly $0.75^\circ$
\cite{Mason02b}, and three mosaic regions, each of size roughly 13
square degrees \cite{Pearson02}. In analyzing such high resolution
data at 30 GHz, great attention must be paid to the contamination by
point sources, but we are confident that this is handled well
\cite{Mason02b}. Data from 2001 roughly doubles the amount, increases
the area covered, and its analysis is currently underway.

\begin{figure}
\includegraphics[width=5.0in]{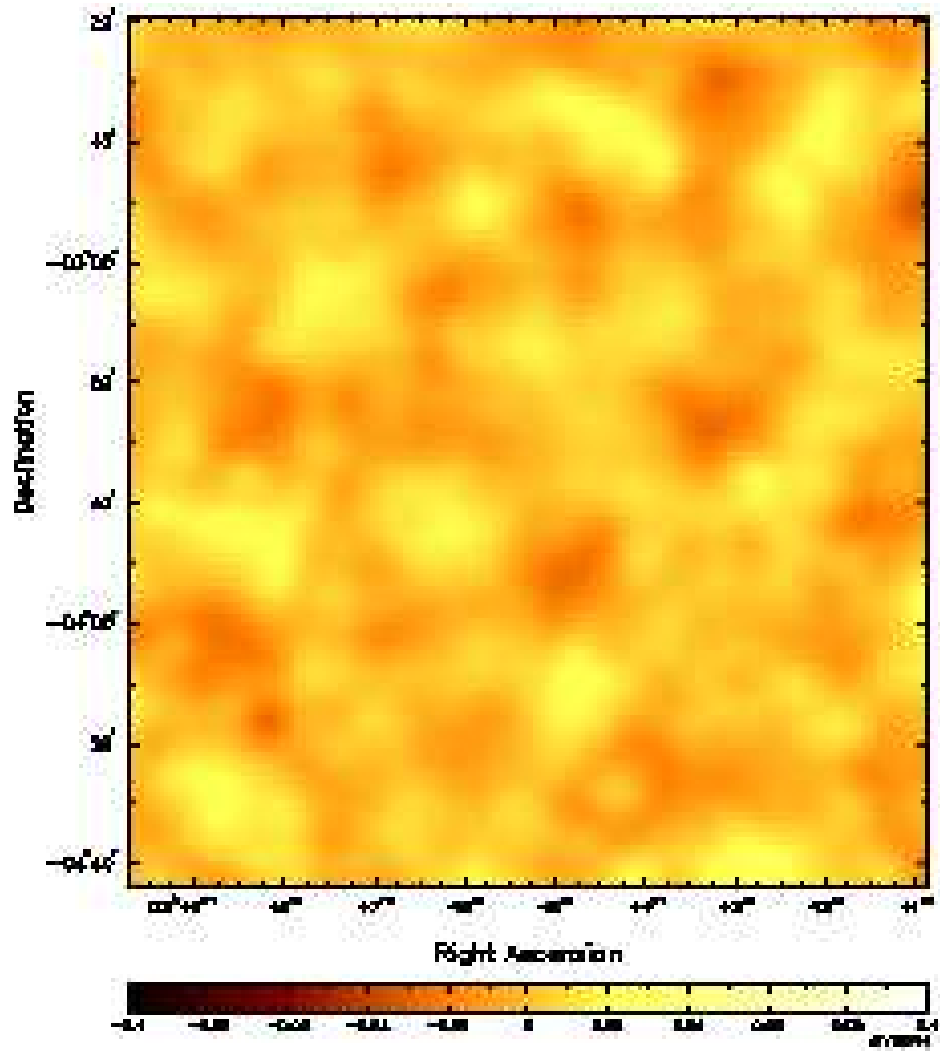}
\caption{The inner 2.5 degrees of one of the three 13 square degree
CBI mosaic fields is shown. The mosaic image is a standard radio
astronomy map, with point sources removed. The regions are smaller
than Boomerang covered, but the resolution is a factor of at least
three better. The mass subtended by the CBI resolution scale ($\sim
4^\prime$) easily encompasses the mass that collapses later in the
universe to generate clusters of galaxies. }
\label{fig:CBImosaicmap}
\end{figure}

\begin{figure}
\includegraphics[width=4.5in]{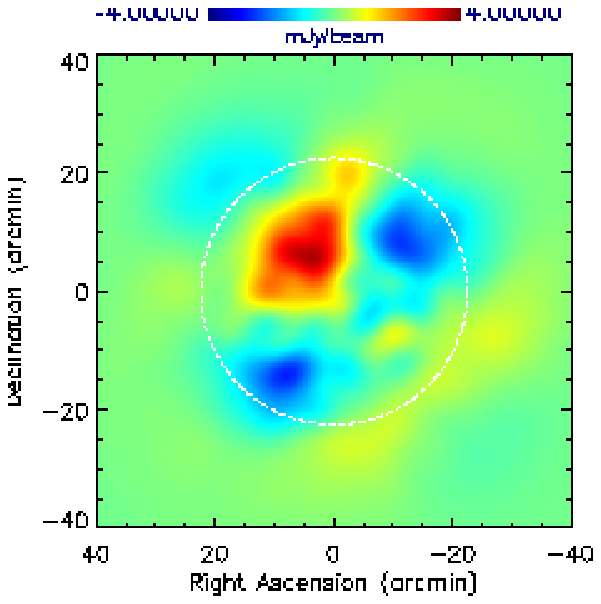}
\caption{An 0.75 degree field-of-view (within the circle)
Wiener-filtered map of one of the three deep CBI fields is shown. Apart
from the primary anisotropy signal, the deep images contain power at
high $\ell$ that might be from the SZ effect in collapsed clusters at
redshift of order 1. We know that the predicted SZ signal is near to that
seen, and is very sensitive to $\sigma_8$ (${\cal C}_\ell^{(SZ)}
\propto \sigma_8^7$). $\sigma_8 \approx 1$ is needed for the SZ effect
as determined by hydro simulations to agree with the data. The primary
CMB data prefer values between 0.8 and 0.9, and so although we do
expect the SZ signal to be lurking within the CBI signal, it may not
be quite as large as the extra power seen in these deep maps.}
\label{fig:CBIdeepmap}
\end{figure}

Fig.~\ref{fig:CBImosaicmap} shows one of the CBI mosaic regions of the
sky and Fig.~\ref{fig:CBIdeepmap} one of the deep regions. 

The VSA (Very Small Array) in Tenerife, also an interferometer,
operating at 30 GHz, covered the $\ell$ range of DASI, and confirmed
the spectrum emerging from the Boomerang, Maxima and DASI data in that
region. The VSA is now observing at longer baselines to increase its 
$\ell$ range. 

\begin{figure}
\includegraphics[width=4.5in]{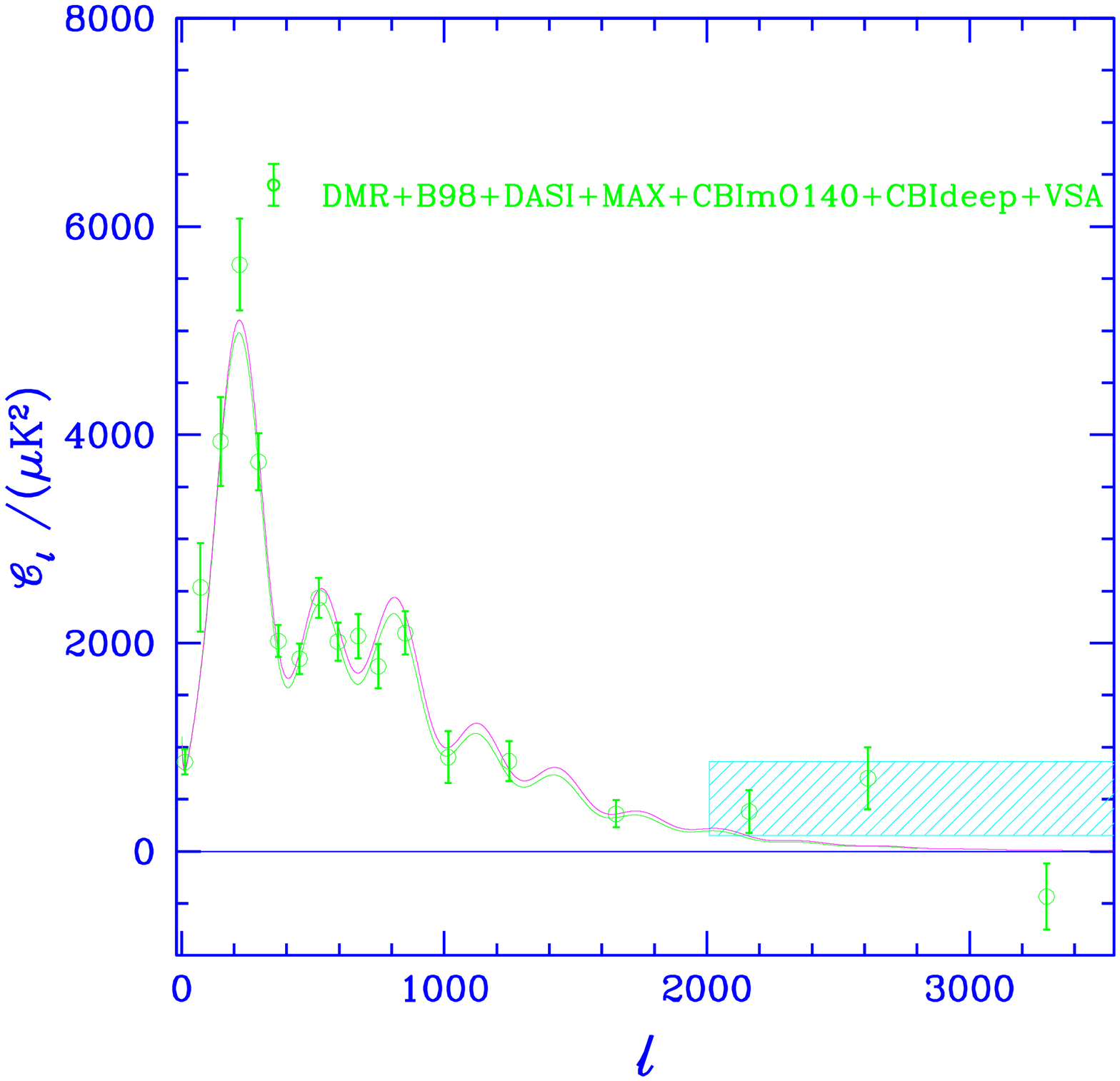}
\caption{The ${\cal C}_\ell$ are defined in terms of CMB temperature
anisotropy multipoles by ${\cal C}_\ell \equiv \ell
(\ell+1)\avrg{\vert (\Delta T/T)_{\ell m}\vert^2}/(2\pi) $.  The
optimal ${\cal C}_\ell$ spectrum corresponds to a maximum-likelihood
fit to the power in bands marginalized over beam and calibration
uncertainties of the various experiments, This one uses ``all-data''
(DMR + Boomerang + Maxima + DASI + CBI mosaic + CBI deep + VSA + TOCO
+ Boomerang-NA + the 4.99 data). A $\Delta \ell =75$ binning was
chosen up to $\sim 800$, going over to the CBI deep binning at large
$\ell$. The $\ell > 2000$ excess found with the deep CBI data is
denoted by the light blue hatched region (95\% confidence limit). Two
best fit models to ``all-data'' are shown. They are both $\Lambda$CDM
models; \eg the upper curve (magenta) has parameters $\{\Omega_{\rm
{tot}}$, $\Omega_\Lambda$, $\Omega_b h^2$, $\Omega_{\rm cdm} h^2$,
$n_s$, $\tau_C\}$=$\{1.0, 0.7, 0.02, 0.14, 0.975, 0\}$. }
\label{fig:CLopt}
\end{figure}

{\bf The Optimal Spectrum, circa Summer 2002:} The power spectrum
shown in Fig.~\ref{fig:CLopt} combines all of the data in a way that
takes all of the uncertainties in each experiment (calibration and
beam) into account. The point at small $\ell$ is dominated by DMR, at
$900 \lta \ell \lta 2000$ by the CBI mosaic data, with that beyond
2000 by the CBI deep data. In between, Boomerang drives the small
error bars, DASI and CBI set the calibration and beam of Boomerang and
give spectra totally compatible with Boomerang. Both VSA and Maxima
are in agreement with this data as well. Further, although the errors
from the experiments before April 2000 are larger, the quite
heterogeneous 4.99+TOCO+Boomerang-NA mix of CMB data is very
consistent with what the newer experiments show. It is an amazing
concordance of data. Accompanying this story is a convergence with
decreasing errors over time on the values of the cosmological
parameters given in Table~\ref{tab:exptparams}. 

{\bf Primary CMB Processes and Soundwave Maps at Decoupling:}
Boomerang, Maxima, DASI, CBI and VSA were designed to measure the {\it
primary} anisotropies of the CMB, those which can be calculated using
linear perturbation theory. What we see in
Figs.~\ref{fig:BOOMmap},~\ref{fig:CBImosaicmap} are, basically, images
of soundwave patterns that existed about 400,000 years after the Big
Bang, when the photons were freed from the plasma. The visually
evident structure on degree scales is even more apparent in the power
spectra of the maps, which show a dominant
(first acoustic) peak, with less prominent subsequent ones detected at
varying levels of statistical significance.

 The images are actually a projected mixture of dominant and
subdominant physical processes through the photon decoupling
"surface", a fuzzy wall at redshift $z_{dec} \approx 1050$, when the
Universe passed from optically thick to thin to Thomson scattering
over a comoving distance $\approx 19 \mpc$. Prior to this, acoustic
wave patterns in the tightly-coupled photon-baryon fluid on scales
below the comoving "sound crossing distance" at decoupling, $\lta 150
\mpc$ (\ie $\lta 150 \kpc$ physical), were viscously damped, strongly
so on scales below the $\approx 10 \mpc$ damping scale. After, photons
freely-streamed along geodesics to us, mapping (through the angular
diameter distance relation) the post-decoupling spatial structures in
the temperature to the angular patterns we observe now as the 
primary anisotropies. The maps are images projected through the
fuzzy decoupling surface of the acoustic waves (photon bunching), the
electron flow (Doppler effect) and the gravitational potential peaks
and troughs ("naive" Sachs-Wolfe effect) back then. Free-streaming
along our (linearly perturbed) past light cone leaves the pattern
largely unaffected, except that temporal evolution in the
gravitational potential wells as the photons propagate through them
leaves a further $\Delta T$ imprint, called the integrated Sachs-Wolfe
effect. Intense theoretical work over three decades has put accurate
calculations of this linear cosmological radiative transfer on a firm
footing, and there are speedy, publicly available and widely used
codes for evaluation of anisotropies in a variety of cosmological
scenarios, ``CMBfast'' and ``CAMB'' \cite{cmbfast}. Extensions to more
cosmological models have been added by a variety of researchers.

Of course there are a number of nonlinear effects that are also
present in the maps. These {\it secondary} anisotropies include
weak-lensing by intervening mass, Thomson-scattering by the
nonlinear flowing gas once it became "reionized" at $z \sim 10$,
the thermal and kinematic SZ effects, and the red-shifted emission
from dusty galaxies. They all leave non-Gaussian imprints on the
CMB sky.

{\bf The Immediate Future:} Results are in and the analysis is
underway for the bolometer single dish ACBAR experiment based at the
South Pole (with about $5^\prime$ resolution, allowing coverage to
$\ell \sim 2000$). Results from two balloon flights, Archeops and
Tophat, are also expected soon.  MINT, a Princeton interferometer,
also has results soon to be released. 

In June 2001, NASA launched the all-sky HEMT-based MAP satellite, with
$12^\prime$ resolution. We expect spectacularly accurate results
covering the $\ell$ range to about 600 to be announced in Jan. 2003,
with higher $\ell$ and smaller errors expected as the observing period 
increases. Eventually four years of data are expected.  

 Further downstream, in 2007, ESA will launch the bolometer+HEMT-based
Planck satellite, with $5^\prime$ resolution.

{\bf Targeting Polarization of the Primary CMB Signal:} The
polarization dependence of Compton scattering induces a well defined
polarization signal emerging from photon decoupling. Given the total ${\cal
C}_\ell$ of Fig.~\ref{fig:CLopt}, we can forecast what the strength 
of that signal and its cross-correlation with the total
anisotropy will be, and which $\ell$ range gives the maximum signal:
$\sim 5 \mu$K over $\ell \sim 400-1600$ is a target for the ``$E$-mode''
that scalar fluctuations give. 

 A great race was on to first detect the $E$-mode.  Experiments range
from many degrees to subarcminute scales. DASI has analyzed 271 days
of polarization data on 2 deep fields ($3.4^\circ$ {\it fov}) and just
announced a $5\sigma$ detection at a level consistent with
inflation-based models \cite{DASIpol02}. Boomerang will fly again, in
December 2002, with polarization-sensitive bolometers of the kind that
will also be used on Planck; as well, MAXIMA will fly again as the
polarization-targeting MAXIPOL; the detectors on CBI have been
reconfigured to target polarization and the CBI is beginning to take
data; MAP can also measure polarization with its HEMTs. Other
experiments, operating or planned include AMiBA, COMPASS, CUPMAP,
PIQUE and its sequel, POLAR, Polarbear, Polatron, QUEST,
Sport/BaRSport, BICEP, among others.  The amplitude of the DASI E
detection is $0.8 \pm 0.3$ of the forecasted amplitude from the total
anisotropy data; the cross-correlation of the polarization with the
total anisotropy has also been detected, with amplitude $0.9 \pm 0.4$
of the forecast. These detections used a broad-band shape covering the
$\ell$ range $\sim 250-750$ derived from the theoretical
forecasts. The forecasts indicate solid detections with enough
well-determined bandpowers for use in cosmological parameter studies
are soon likely.

We cannot yet forecast the strength of the ``$B$-mode'' signal induced
by gravity waves, since there is as yet no evidence for or against
them in the data. However, the amplitude would be very small indeed
even at $\ell \sim 100$. Nonetheless there are experiments such as
BICEP (Caltech) being planned to go after polarization in these low $\ell$
ranges.

{\bf Targeting Secondary Anisotropies:} SZ anisotropies have been
probed by single dishes, the OVRO and BIMA mm arrays, and the Ryle
interferometer. Detections of individual clusters are now routine. The
power at $\ell > 2000$ seen in the CBI deep data
(Fig.~\ref{fig:CLopt}) \cite{Mason02b} and in the BIMA data at $\ell
\sim 6000$ \cite{BIMA02} may be due to the SZ effect in ambient
fields, \eg \cite{Bond02}. A number of planned HEMT-based
interferometers are being built with this ambient effect as a target:
CARMA (OVRO+BIMA together), the SZA (Chicago, to be incorporated in
CARMA), AMI (Britain, including the Ryle telescope), and AMiBA
(Taiwan). Bolometer-based experiments will also be used to probe the
SZ effect, including the CSO (Caltech submm observatory, a 10m dish)
with BOLOCAM on Mauna Kea, ACBAR at the South Pole and the LMT (large
mm telescope, with a 50 metre dish) in Mexico.  As well, APEX, a 12 m
German single dish, Kobyama, a 10 m Japanese single dish, and the 100m
Green Bank telescope can all be brought to bear down on the SZ
sky. Very large bolometer arrays with thousands of elements are
planned: the South Pole Telescope (SPT, Chicago) and the Atacama
Cosmology Telescope (ACT, Princeton), with resolution below
$2^\prime$, will be powerful probes of the SZ effect as well as of
primary anisotropies.

Anisotropies from dust emission from high redshift galaxies are being
targeted by the JCMT with the SCUBA bolometer array, the OVRO mm
interferometer, the CSO, the SMA (submm array) on Mauna Kea, the LMT,
the ambitious US/ESO ALMA mm array in Chile, the LDB BLAST, and ESA's
Herschel satellite. About $50\%$ of the submm background has so far been
identified with sources that SCUBA has found.

{\bf The CMB Analysis Pipeline:} Analyzing Boomerang and other
single-dish experiments involves a pipeline, reviewed in \cite{bc01},
that takes the timestream in each of the bolometer channels coming
from the balloon plus information on where it is pointing and turns it
into spatial maps for each frequency characterized by average
temperature fluctuation values in each pixel (Fig.~\ref{fig:BOOMmap})
and a pixel-pixel correlation matrix characterizing the noise. From
this, various statistical quantities are derived, in particular the
temperature power spectrum as a function of multipole, grouped into
bands, and band-band error matrices which together determine the full
likelihood distribution of the
bandpowers~\cite{BJK98,BJK2000}. Fundamental to the first step is the
extraction of the sky signal from the noise, using the only
information we have, the pointing matrix mapping a bit in time onto a
pixel position on the sky. In the April 2001 analysis of Boomerang,
and subsequent work, powerful use of Monte Carlo simulations was made
to evaluate the power spectrum and other statistical indicators in
maps with many more pixels than was possible with conventional matrix
methods for estimating power spectra \cite{Netterfield01}.

For interferometer experiments, the basic data are visibilities as a
function of baseline and frequency, with contributions from random
detector noise as well as from the sky signals. As mentioned above, a
baseline is a direct probe of a given angular wavenumber vector on the
sky, hence suggests we should make ``generalized maps'' in ``momentum
space'' (\ie Fourier transform space) rather than in position space,
as for Boomerang. A major advance was made in \cite{Myers02} to deal
with the large volume of interferometer data that we got with CBI,
especially for the mosaics with their large number of overlapping
fields.  We ``optimally'' compressed the $> {\cal O}(10^5)$ visibility
measurements of each field into a $ < {\cal O}(10^4)$ coarse grained
lattice in momentum space, and used the information in that
``generalized pixel'' basis to estimate the power spectrum and
statistical distribution of the signals.

There is generally another step in between the maps and the final
power spectra, namely separating the multifrequency spatial maps into
the physical components on the sky: the primary CMB, the thermal and
kinematic Sunyaev-Zeldovich effects, the dust, synchrotron and
bremsstrahlung Galactic signals, the extragalactic radio and
submillimetre sources. The strong agreement among the Boomerang maps
indicates that to first order we can ignore this step, but it has to
be taken into account as the precision increases.  

Because of the 1 cm observing wavelength of CBI and its resolution,
the contribution from extragalactic radio sources is significant.  We
project out of the data sets known point sources when estimating the
primary anisotropy spectrum by using a number of constraint
matrices. The positions are obtained from the (1.4 GHz) NVSS
catalog. When projecting out the sources we use large amplitudes which
effectively marginalize over all affected modes. This insures
robustness with respect to errors in the assumed fluxes of the
sources. The residual contribution of sources below our known-source
cutoff is treated as a white noise background with an amplitude (and
error) estimated as well from the NVSS database \cite{Mason02b}.

{\bf The CMB Statistical Distributions are Nearly Gaussian:}
The primary CMB fluctuations are quite Gaussian, according to COBE,
Maxima, and now Boomerang and CBI analyses. Analysis of data like that
in the Fig.~\ref{fig:BOOMmap} map show a one-point distribution of
temperature anisotropy values that is well fit by a Gaussian
distribution.  Higher order (concentration) statistics (3,4-point
functions, \etc) tell us of non-Gaussian aspects, necessarily expected
from the Galactic foreground and extragalactic source signals, but
possible even in the early Universe fluctuations. For example, though
non-Gaussianity occurs only in the more baroque inflation models of
quantum noise, it is a necessary outcome of defect-driven models of
structure formation. (Peaks compatible with Fig.~\ref{fig:CLopt} do
not appear in non-baroque defect models, which now appear highly
unlikely.)  There is currently no evidence for a breakdown of the
Gaussianity in the 150 GHz maps as long as one does not include
regions near the Galactic plane in the analysis. We have also found
that the one-point distribution of the CBI data is also compatible
with a Gaussian. However, since we know non-Gaussianity is necessarily
there at some level, more exploration is needed.

Though great strides have been made in the analysis of the
Boomerang-style and CBI-style experiments, there is intense effort
worldwide developing fast and accurate algorithms to deal with the
looming megapixel datasets of LDBs and the satellites. Dealing more
effectively with the various component signals and the statistical
distribution of the errors resulting from the component separation is
a high priority.

\section{Cosmic Parameter Estimation} \label{sec:params}

{\bf Parameters of Structure Formation:} Following \cite{capp00}, we
adopt a restricted set of 8 cosmological parameters, augmenting the
basic 7 used in \cite{lange00,jaffe00,Netterfield01,Sievers02},
$\{\Omega_\Lambda,\Omega_{k},\omega_b,\omega_{cdm}, n_s,\tau_C,
\sigma_8\}$, by one. The vacuum or dark energy encoded in the
cosmological constant $\Omega_\Lambda$ is reinterpreted as $\Omega_Q$,
the energy in a scalar field $Q$ which dominates at late times, which,
unlike $\Lambda$, could have complex dynamics associated with it. $Q$
is now often termed a quintessence field.  One popular phenomenology
is to add one more parameter, $w_Q = p_Q /\rho_Q$, where $p_Q$ and
$\rho_Q$ are the pressure and density of the $Q$-field, related to its
kinetic and potential energy by $\rho_Q = \dot{Q}^2/2+(\nabla
Q)^2/2+V(Q)$, $ p_Q = \dot{Q}^2/2 -(\nabla Q)^2/6-V(Q)$. Thus $w_Q=-1$
for the cosmological constant. Spatial fluctuations of $Q$ are
expected to leave a direct imprint on the CMB for small $\ell$. This
will depend in detail upon the specific model for $Q$. We ignore this
complication here, but caution that using DMR data which is sensitive
to low $\ell$ behaviour in conjunction with the rest will give
somewhat misleading results. To be self consistent, a model must be
complete: \eg even the ludicrous models with constant $w_Q $ would
have necessary fluctuations to take into account. As well, as long as
$w_Q$ is not exactly $-1$, it will vary with time, but the data will
have to improve for there to be sensitivity to this, and for now we
can just interpret $w_Q$ as an appropriate time-average of the
equation of state. The curvature energy $\Omega_k \equiv
1-\Omega_{tot}$ also can dominate at late times, as well as affecting
the geometry.

We use only 2 parameters to characterize the early universe primordial
power spectrum of gravitational potential fluctuations $\Phi$, one
giving the overall power spectrum amplitude ${\cal P}_{\Phi}(k_n)$,
and one defining the shape, a spectral tilt $n_s (k_n) \equiv 1+d\ln
{\cal P}_{\Phi}/d \ln k$, at some (comoving) normalization wavenumber
$k_n$. We really need another 2, ${\cal P}_{GW}(k_n)$ and $n_t(k_n)$,
associated with the gravitational wave component. In inflation, the
amplitude ratio is related to $n_t$ to lowest order, with ${\cal
O}(n_s-n_t)$ corrections at higher order, \eg \cite{bh95}. There are
also useful limiting cases for the $n_s-n_t$ relation. However, as one
allows the baroqueness of the inflation models to increase, one can
entertain essentially any power spectrum (fully $k$-dependent $n_s(k)$
and $n_t(k)$) if one is artful enough in designing inflaton potential
surfaces. Actually $n_t(k)$ does not have as much freedom as $n_s(k)$
in inflation. For example, it is very difficult to get $n_t(k)$ to be
positive.  As well, one can have more types of modes present, \eg
scalar isocurvature modes (${\cal P}_{is}(k_n),n_{is}(k)$) in addition
to, or in place of, the scalar curvature modes (${\cal
P}_{\Phi}(k_n),n_{s}(k)$). However, our philosophy is consider minimal
models first, then see how progressive relaxation of the constraints
on the inflation models, at the expense of increasing baroqueness,
causes the parameter errors to open up. For example, with COBE-DMR and
Boomerang, we can probe the GW contribution, but the data are not
powerful enough to determine much. Planck can in principle probe the
gravity wave contribution reasonably well.

We use another 2 parameters to characterize the transport of the
radiation through the era of photon decoupling, which is sensitive
to the physical density of the various species of particles
present then, $\omega_j \equiv \Omega_j {\rm h}^2$. We really need
4: $\omega_b$ for the baryons, $\omega_{cdm}$ for the cold dark
matter, $\omega_{hdm}$ for the hot dark matter (massive but light
neutrinos), and $\omega_{er}$ for the relativistic particles
present at that time (photons, very light neutrinos, and possibly
weakly interacting products of late time particle decays). For
simplicity, though, we restrict ourselves to the conventional 3
species of  relativistic neutrinos plus photons, with
$\omega_{er}$ therefore fixed by the CMB temperature and the
relationship between the neutrino and photon temperatures
determined by the extra photon entropy accompanying $e^+ e^- $
annihilation. Of particular importance for the pattern of the
radiation is the (comoving) distance sound can have influenced by
recombination (at redshift $z_{dec}= a_{dec}^{-1}-1$), 
\begin{equation} 
r_s =
6000/\sqrt{3} \mpc \int_{0}^{\sqrt{a_{dec}}} (\omega_m + \omega_{er}
a^{-1})^{-1/2} (1+ \omega_b a/(4\omega_\gamma /3))^{-1/2}\
d\sqrt{a}, 
\end{equation}
where $\omega_\gamma = 2.46 \times 10^{-5}$ is the
photon density, $\omega_{er} = 1.68 \omega_\gamma$ for 3 species
of massless neutrinos and $\omega_m \equiv
\omega_{hdm}+\omega_{cdm}+\omega_b$.

 The angular diameter distance relation maps spatial structure at
 photon decoupling perpendicular to the line-of-sight with transverse
 wavenumber $k_\perp$ to angular structure, through $\ell = {\cal
 R}_{dec} k_\perp$. In terms of the comoving distance to photon
 decoupling (recombination), $\chi_{dec} $, and the curvature scale
 $d_k$, ${\cal
 R}_{dec}$ is given by 
\begin{eqnarray} 
&& {\cal R}_{dec} =\{d_k {\rm sinh} (\chi_{dec}/d_k), \chi_{dec}, d_k
{\rm sin} (\chi_{dec}/d_k)\}, \ {\rm where} \ d_k =3000
|\omega_k|^{-1/2} \mpc, \nonumber \\ && {\rm and} \ \chi_{dec} = 6000
\mpc \int_{\sqrt{a_{dec}}}^{1} (\omega_m + \omega_Q a^{-6w_Q}
+\omega_k a)^{-1/2}\ d\sqrt{a} \, .
\end{eqnarray}
The 3 cases are for negative, zero and positive mean curvature. Thus
the mapping depends upon $\omega_{k}$, $\omega_Q$ and $w_Q$ as well as
on $\omega_m$. The location of the acoustic peaks $\ell_{pk,j}$ is
proportional to the ratio of ${\cal R}_{dec}$ to $r_s$, hence depends
upon $\omega_b$ through the sound speed as well.  Thus $\ell_{pk,j}$
defines a functional relationship among these parameters, a {\it
degeneracy} \cite{degeneracies} that would be exact except for the
integrated Sachs-Wolfe effect, associated with the change of $\Phi$
with time if $\Omega_Q$ or $\Omega_k$ is nonzero. (If $\dot{\Phi}$
vanishes, the energy of photons coming into potential wells is the
same as that coming out, and there is no net impact of the rippled
light cone upon the observed $\Delta T$.)

 Our 7th parameter is an astrophysical one, the Compton "optical
depth" $\tau_C$ from a reionization redshift $z_{reh}$ to the
present. It lowers ${\cal C}_\ell$ by $\exp(-2\tau_C)$ at $\ell$'s in
the Boomerang/CBI regime. For typical models of hierarchical structure
formation, we expect $\tau_C \lta 0.2$. It is partly degenerate with
$\sigma_8$ and cannot quite be determined at this precision by CMB
data now.

The LSS also depends upon our parameter set: an important combination
is the wavenumber of the horizon when the energy density in
relativistic particles equals the energy density in nonrelativistic
particles: $k_{Heq}^{-1} \approx 5 \Gamma^{-1} \hmpc$, where $\Gamma
\approx \Omega_m {\rm h} \Omega_{er}^{-1/2}$. Instead of ${\cal
P}_\Phi (k_n)$ for the amplitude parameter, we often use ${\cal
C}_{10}$ at $\ell =10$ for CMB only, and $\sigma_8^2$ when LSS is
added. When LSS is considered in this paper, it refers to constraints
on $\Gamma + (n_s-1)/2$ and $\ln \sigma_8^2$ that are obtained by
comparison with the data on galaxy clustering, cluster abundances and
from weak lensing~ \cite{Bond02,lange00,bj99}. At the current time,
the constraints from $\sigma_8$ from lensing and cluster abundances
are stronger thoes from $\Gamma$, although, with the wealth of data
emerging from the Sloan Digital Sky Survey and the 2dF redshift
survey, shape should soon deliver more powerful information than
overall amplitude. However, in the future, weak lensing will allow
amplitude and shape to be simultaneously constrained without the
uncertainties associated with the biasing of the galaxy distribution
{\it wrt} the mass that the redshift surveys must deal with.

When we allow for freedom in $\omega_{er}$, the abundance of
primordial helium, tilts of tilts ($dn_{\{s,is,t\}}(k_n)/d\ln k,
...$) for 3 types of perturbations, the parameter count would be
17, and many more if we open up full theoretical freedom in
spectral shapes. However, as we shall see, as of now only 4
combinations can be determined with 10\% accuracy with the CMB.
Thus choosing 8 is adequate for the present; 7 of these are
discretely sampled\cite{database}, with generous boundaries,
though for drawing cosmological conclusions we adopt a weak prior
probability on the Hubble parameter and age: we restrict ${\rm h}
$ to lie in the 0.45 to 0.9 range, and the age to be above 10 Gyr.

\begin{figure}
\includegraphics[width=4.5in]{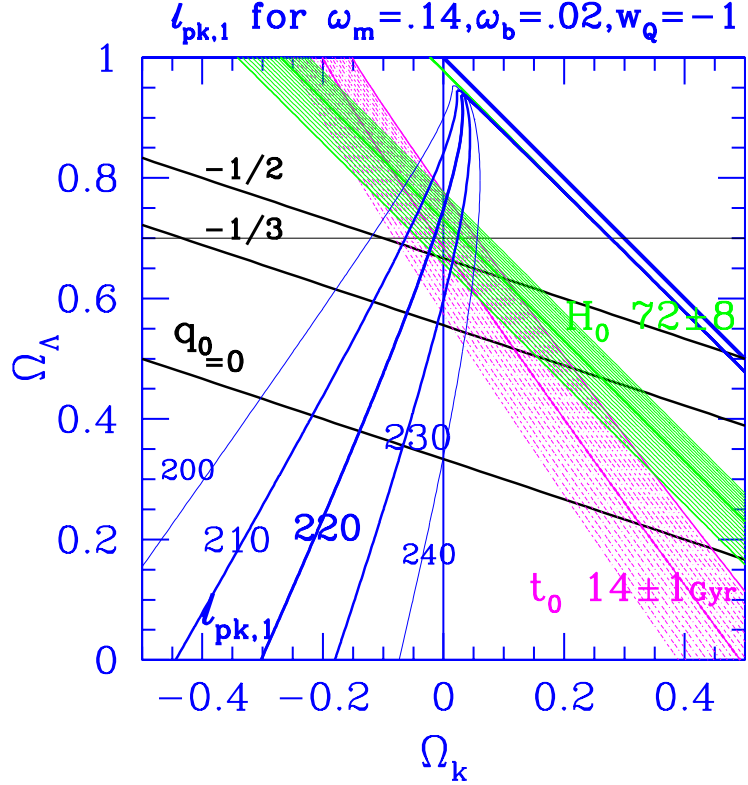}
\caption{ This shows lines of constant $\ell_{pk,1}$ in the
$\Omega_k$--$\Omega_Q$ plane (assuming $w_Q$=$-1$, \ie a cosmological
constant) for the $\{ \omega_m ,\omega_b \}$ shown, near their most
probable values. The data give $\ell_{pk,1} = 222 \pm 3$. The higher
peaks and dips have similar curves, scaled about the probable values
listed above.  Lowering $\omega_b$ increases the sound speed,
decreasing $\ell_{pk,j}$, and varying $\omega_m$ also shifts it. The
$0.64 < {\rm h}< 0.82 $ (heavier shading, $H_0$) and $13< $ age $< 15$
(lighter shading, $t_0$) ranges and decelerations $q_0=0,-1/3,-1/2$
are also noted.  The sweeping back of the $\ell_{pk,j}$ curves into
the closed models as $\Omega_\Lambda$ is lowered shows that even if
$\Omega_{tot}$=1, the phase space results in a 1D projection onto the
$\Omega_{tot}$ axis that would be skewed to $\Omega_{tot}>1$. This
plot explains much of the structure in the probability contour maps
derived from the data, Fig.~\ref{fig:OmkOLb}.  }
\label{fig:OmkOL}
\end{figure}

\begin{figure}
\includegraphics[width=4.5in]{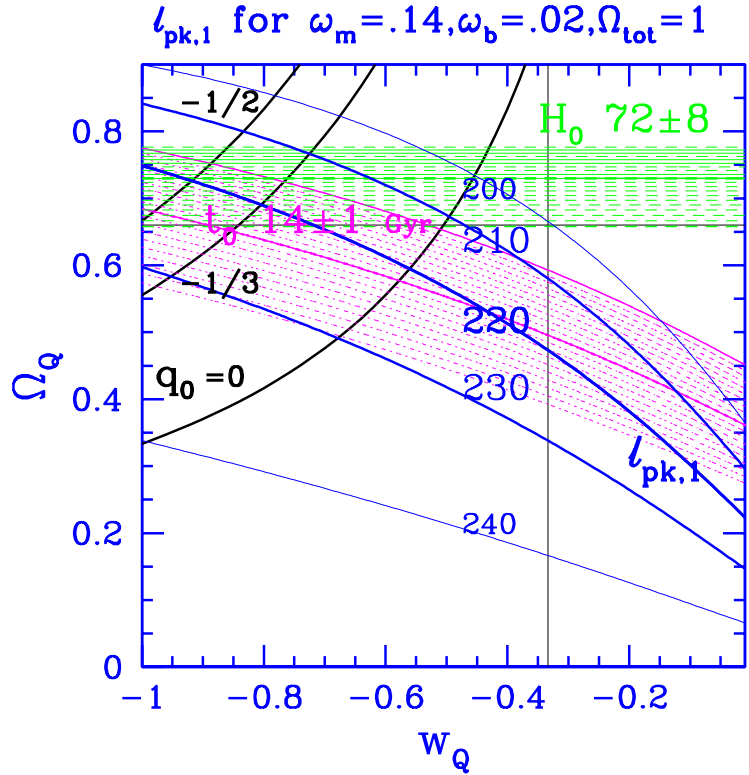}
\caption{ Lines of constant $\ell_{pk,1}$ in the $w_Q$--$\Omega_Q$
quintessence plane (with $\Omega_{tot}$=1) are shown for the most
probable values of $\{ \omega_m ,\omega_b \}$. Lines of constant
deceleration parameter $q_0 = (\Omega_m +(1+3w_Q)\Omega_Q )/2$, $H_0$
and age $t_0$ in the ranges indicated are also shown.  }
\label{fig:wQOQ}
\end{figure}

{\bf Peaks, Dips and $\Omega_{tot}$, $\Omega_Q$ and $w_Q$:} For given
$\omega_m$ and $\omega_b$, we show the lines of constant
$\ell_{pk,j}\propto {\cal R}_{dec}/r_s$ in the
$\Omega_{k}$--$\Omega_Q$ plane for $w_Q$=$-1$ in Fig.~\ref{fig:OmkOL},
and in the $w_Q$--$\Omega_Q$ plane for $\Omega_{tot}$=1 in
Fig.~\ref{fig:wQOQ}, using the formulas given above and in
\cite{degeneracies}.  

Our current best estimate \cite{Sievers02} of
the peak locations $\ell_{pk,j}$, using all current CMB data and the
flat+wk-$h$+LSS prior, are $222 \pm 3$, $537 \pm 6$, $823 \pm 45$,
$1138 \pm 45$, $1437 \pm 59$, obtained by forming $\exp<\ln
\ell_{pk,j}>$, where the average and variance of $\ln \ell_{pk,j}$ are
determined by integrating over the probability-weighted database
described above. The interleaving dips are at $411 \pm 5$, $682 \pm
48$, $1023 \pm 44$, $1319 \pm 51$, $1653 \pm 48$. With just the data
prior to April 2000, the first peak value was $224\pm 25$, showing how
it has localized.  A quadratic fit sliding over the data can be
used to estimate the peak and dip positions in a model-independent
way. It gives numbers in good accord with those given here
\cite{deBernardis01,Pearson02}, but of course with larger error bars,
in some cases only one-sigma detections for the higher peaks and dips.

The critical spatial scale determining the positions of the peaks is
$r_s$, found by averaging over the model space probabilities to be
$145 \pm 2 \mpc$ comoving, thus about $140 \kpc$ as the physical sound
horizon at decoupling.  Converting peaks in $k$-space into peaks in
$\ell$-space is obscured by projection effects over the finite width
of decoupling and the influence of sources other than sound
oscillations such as the Doppler term. The conversion into peak
locations in ${ \cal C}_\ell$ gives $\ell_{pk,j} \sim f_j j \pi {\cal
R}_{dec}/r_s$, where the numerically estimated $f_j$ factor is
$\approx 0.75$ for the first peak, approaching unity for higher
ones. Dip locations are determined by replacing $j$ by $j+1/2$. These
numbers accord reasonably well with the ensemble-averaged
$\ell_{pk/dip, j}$ given above.

The strength of the overall decline due to shear viscosity and the
finite width of the region over which hydrogen recombination occurs
can also be estimated, $R_D = 10 \pm 3 \mpc$, \ie $10 \kpc$ back then,
corresponding to an angular damping scale $\ell_D = 1358 \pm 22$.

\begin{figure}
\includegraphics[width=4.0in]{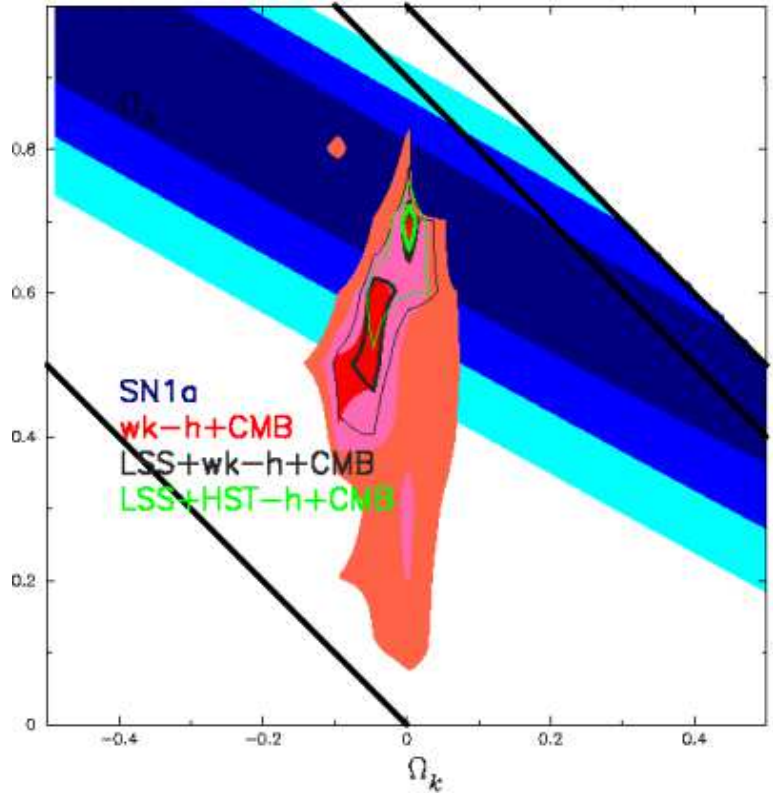}
\caption{ 1,2,3-sigma likelihood contour shadings for ``all-data'' and
the weak-H+age prior probability in the $\Omega_k-\Omega_Q$ plane. The
first interior lines are the 1,2-sigma ones when the LSS constraint is
added, the most interior are the contours when the Hubble key project
constraint is applied.  The supernova contour shadings \cite{SN} are
also plotted.  Note that the contours are near the $\Omega_{k}=0$
line, but also follow a weighted average of the $\ell_{pk,1}\sim 220$
lines. This approximate degeneracy implies $\Omega_Q$ is poorly
constrained for CMB-only, but it is broken when LSS is added, giving a
solid SN1-independent $\Omega_Q$ "detection".  When the Hubble key
project constraint on $H_0$ is added, partial breaking of this
degeneracy occurs as well, as is evident from Fig.~\ref{fig:OmkOL},
and from this figure. }
\label{fig:OmkOLb}
\end{figure}

\begin{figure}
\includegraphics[width=4.0in]{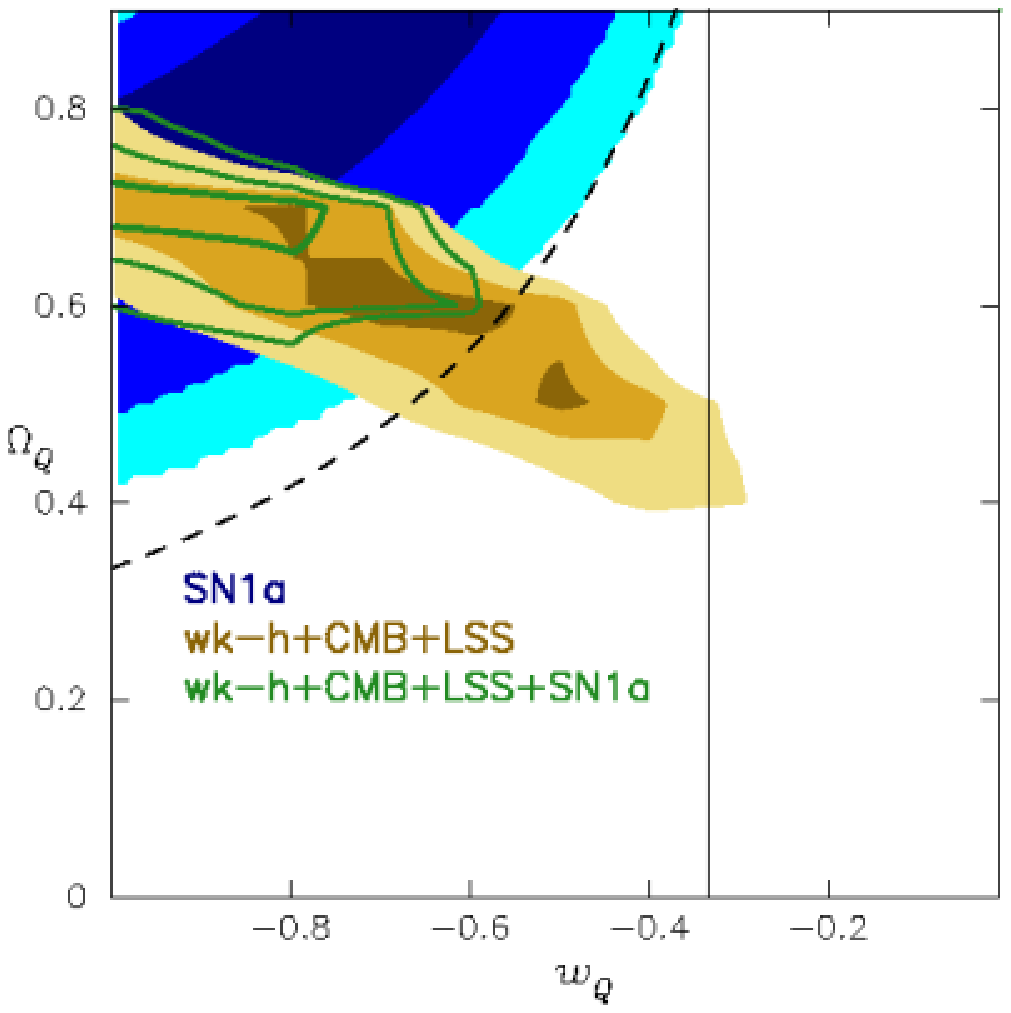}
\caption{1,2,3-sigma likelihood contour regions for ``all-data'' with
the weak-H+age prior probability, $\Omega_{tot}$=1, and the LSS
prior. 1,2,3-sigma SN1 contour regions~\protect\cite{perlmutter99} at
upper right are rather similar to the constant deceleration lines.
The $q_0=0$ line is the dashed one. The combined CMB+LSS+SN1 contours
are shown as lines. The constraint from CMB on $\Omega_Q$ is
reasonably good if a strong $H_0$ prior (Hubble key project result) is
applied, but gets significantly more localized when LSS is added. In
neither case is $w_Q$ well determined, and $w_Q$ localization with
CMB+LSS+SN1 is mainly because of SN1. This implies the current limits
are not that sensitive to the details of how quintessence impacts low
$\ell$ in the DMR regime, which is an uncertainty in the theory.}
\label{fig:wQOQb}
\end{figure}

The constant $\ell_{pk,j}$ lines look rather similar to the contours
in Figs.~\ref{fig:OmkOLb},~\ref{fig:wQOQb}, showing that the
${\cal R}_{dec}/r_s$ degeneracy plays a large role in determining the
contours.  The figures also show how adding other cosmological
information such as $H_0$ estimation can break the degeneracies. The
contours hug the $\Omega_{k}=0$ line more closely than the allowed
$\ell_{pk,j}$ band does for the maximum probability values of
$\omega_m$ and $\omega_b$, because of the shift in the allowed
$\ell_{pk,j}$ band as $\omega_m$ and $\omega_b$ vary in this
plane. The $\omega_b$ dependence in $r_s$ would lead to a degeneracy
with other parameters in terms of peak/dip positions. However,
relative peak/dip heights are extremely significant for parameter
estimation as well, and this breaks the degeneracy. For example,
increasing $\omega_b$ beyond the nucleosynthesis (and CMB) estimate
leads to a diminished height for the second peak that is not in accord
with the data.

{\bf Marginalized Estimates of our Basic 8 Parameters:}
Table~\ref{tab:exptparams} shows there are strong detections with only
the CMB data for $\Omega_{tot}$, $\omega_b$ and $n_s$ in the minimal
inflation-based 8 parameter set. The ranges quoted are Bayesian 50\%
values and the errors are 1-sigma, obtained after projecting
(marginalizing) over all other parameters. With ``all-data'',
$\omega_{cdm}$ begins to localize, but more so when LSS information is
added. Indeed, even with just the COBE-DMR+LSS data, $\omega_{cdm}$ is
already localized. That $\Omega_Q$ is not well determined is a
manifestation of the $\Omega_{k}$--$\Omega_Q$ near-degeneracy
discussed above, which is broken when LSS is added because the
CMB-normalized $\sigma_8$ is quite different for open cf. pure
$Q$-models. Supernova at high redshift give complementary information
to the CMB, but with CMB+LSS (and the inflation-based paradigm) we do
not need it: the CMB+SN1 and CMB+LSS numbers are quite compatible. In
our space, the Hubble parameter, ${\rm h}= (\sum_j (\Omega_j{\rm h}^2
))^{1/2}$, and the age of the Universe, $t_0$, are derived functions
of the $\Omega_j{\rm h}^2$: representative values are given in the
Table caption. CMB+LSS does not currently give a useful constraint on
$w_Q$, though $w_Q \lta -0.7$ with SN1. The values do not change very
much if rather than the weak prior on ${\rm h}$, we use $0.72 \pm
0.08$, the estimate from the Hubble key project
\cite{Freedman}. Indeed just the CMB data plus this restricted range
for ${\rm h}$ and the restriction to $\Omega_{tot}=1$ results in a
strong detection of $\Omega_Q$. Allowing for a neutrino mass
\cite{mnu} changes the value of $\Omega_Q$ downward as the mass
increases, but not so much as to make it unnecessary.

\begin{table}
\caption{Cosmological parameter values and their 1-sigma errors are
shown, determined after marginalizing over the other $6$ cosmological
and the various experimental parameters, for ``all-data'' and
the weak prior ($0.45 \le {\rm h} \le 0.9$, age $> 10$ Gyr). The LSS
prior was also designed to be weak. The detections are clearly very
stable if extra "prior" probabilities for LSS and SN1 are
included. Similar tables are given in \protect\cite{Sievers02}. If
$\Omega_{tot}$ is varied, but $w_Q=-1$, parameters derived from our
basic 8 come out to be: age=$15.2\pm 1.3$ Gyr, ${\rm h}=0.55 \pm
0.08$, $\Omega_m=0.46\pm .11$, $\Omega_b=0.070 \pm .02$. Restriction
to $\Omega_{tot}=1$ and $w_Q=-1$ yields: age=$14.1\pm 0.6$ Gyr, ${\rm
h}=0.65\pm .05$, $\Omega_m=0.34\pm .05$, $\Omega_b=0.05 \pm .006$;
allowing $w_Q$ to vary yields quite similar results. }
\label{tab:exptparams}
\begin{tabular}{|l|llll|}
\hline
   & cmb & +LSS & +SN1 &  +SN1+LSS \\
\hline
 & $\Omega_{tot}$ & variable & $w_Q=-1$ & CASE  \\
\hline $\Omega_{tot}$           & $1.03^{+.04}_{-.05}$ &
$1.03^{+.03}_{-.04}$ & $1.01^{+.07}_{-.03}$ & $1.00^{+.07}_{-.03}$ \\
$\Omega_b{\rm h}^2$             & $.022^{+.004}_{-.002}$ &
$.022^{+.004}_{-.002}$ & $.023^{+.005}_{-.003}$ &
$.023^{+.004}_{-.003}$  \\
$\Omega_{cdm}{\rm h}^2$ & $.13^{+.03}_{-.03}$ &
$.12^{+.02}_{-.03}$ & $.11^{+.03}_{-.03}$ & $.11^{+.03}_{-.03}$  \\
$n_s$            & $0.94^{+.11}_{-.04}$ & $0.93^{+.11}_{-.04}$ &
$0.97^{+.15}_{-.06}$ & $0.99^{+.14}_{-.07}$ \\
$\Omega_{Q}$ & $0.53^{+.14}_{-.13}$ & $0.57^{+.13}_{-.08}$ &
$0.70^{+.07}_{-.08}$ & $0.70^{+.06}_{-.07}$  \\
\hline
 & $\Omega_{tot}$  & =1 & $w_Q=-1$ & CASE  \\
\hline $\Omega_b{\rm h}^2$& $.021^{+.003}_{-.002}$ &
$.022^{+.003}_{-.002}$ & $.022^{+.003}_{-.002}$ &
$.022^{+.003}_{-.002}$ \\
$\Omega_{cdm}{\rm h}^2$& $.14^{+.03}_{-.02}$ & $.12^{+.01}_{-.01}$
& $.12^{+.01}_{-.01}$ & $.12^{+.01}_{-.01}$\\
$n_s$ & $0.93^{+.05}_{-.03}$ & $0.95^{+.09}_{-.04}$ &
$0.95^{+.09}_{-.04}$ & $0.97^{+.08}_{-.05}$  \\
$\Omega_{Q}$& $0.61^{+.09}_{-.38}$ & $0.66^{+.05}_{-.06}$ &
$0.68^{+.03}_{-.05}$ & $0.68^{+.03}_{-.05}$ \\
\hline
 & $\Omega_{tot}$  & =1 &  $w_Q$ variable & CASE  \\
\hline $\Omega_b{\rm h}^2$& $.021^{+.003}_{-.002}$ &
$.022^{+.003}_{-.002}$ & $.022^{+.003}_{-.002}$ &
$.022^{+.003}_{-.002}$ \\
$\Omega_{cdm}{\rm h}^2$& $.14^{+.03}_{-.02}$ & $.12^{+.01}_{-.01}$
& $.12^{+.01}_{-.01}$ & $.12^{+.01}_{-.01}$\\
$n_s$ & $0.95^{+.05}_{-.04}$ & $0.96^{+.07}_{-.04}$ &
$0.96^{+.07}_{-.04}$ & $0.98^{+.07}_{-.05}$  \\
$\Omega_{Q}$& $0.54^{+.12}_{-.28}$ & $0.61^{+.07}_{-.07}$ &
$0.69^{+.03}_{-.05}$ & $0.69^{+.03}_{-.05}$ \\
$w_{Q}$ (95\%)& $< -0.43$ & $<-0.46$ & $< -0.71$ & $<-0.71$  \\
\hline
\end{tabular}
\end{table}

{\bf The Future, Forecasts for Parameter Eigenmodes:} We can also
forecast dramatically improved precision with future LDBs,
ground-based single dishes and interferometers, MAP and Planck.
Because there are correlations among the physical variables we wish to
determine, including a number of near-degeneracies beyond that for
$\Omega_{k}$--$\Omega_Q$ \cite{degeneracies}, it is useful to
disentangle them, by making combinations which diagonalize the error
correlation matrix, "parameter eigenmodes" \cite{bh95,degeneracies}.
For this exercise, we will add $\omega_{hdm}$ and $n_t$ to our
parameter mix, but set $w_Q$=$-1$, making 9. (The ratio ${\cal
P}_{GW}(k_n)/{\cal P}_\Phi (k_n)$ is treated as fixed by $n_t$, a
reasonably accurate inflation theory result.) The forecast for
Boomerang based on the 800 sq. deg. patch with four 150 GHz bolometers
used is 4 out of 9 linear combinations should be determined to $\pm
0.1$ accuracy. This is indeed what was obtained in the full analysis
of CMB only for Boomerang+DMR. The situation improves for the
satellite experiments: for MAP, with 2 years of data, we forecast 6/9
combos to $\pm 0.1$ accuracy, 3/9 to $\pm 0.01$ accuracy; for Planck,
7/9 to $\pm 0.1$ accuracy, 5/9 to $\pm 0.01$ accuracy. While we can
expect systematic errors to loom as the real arbiter of accuracy, the
clear forecast is for a very rosy decade of high precision CMB
cosmology that we are now fully into.


\def\prd{{Phys.~Rev.~D}}
\def\prl{{Phys.~Rev.~Lett.}}
\def\apj{{Ap.~J.}}
\def\apjl{{Ap.~J.~Lett.}}
\def\apjsuppl{{Ap.~J.~Supp.}}
\def\mnras{{M.N.R.A.S.}}


\begin{thebibliography}{99}
\bibitem{database}The specific discrete parameter values used for the
${\cal C}_\ell$-database in this analysis were: ($\Omega_Q =$
0,.1,.2,.3,.4,.5,.6,.7,.8,.9,1.0,1.1), ($\Omega_k =$
.9,.7,.5,.3,.2,.15,.1,.05,0,-.05,-.1,-.15,-.2,-.3,-.5) \& ($\tau_c
=$0, .025, .05, .075, .1, .15, .2, .3, .5) when $w_Q=$ --1; ($\Omega_Q
=$ 0,.1,.2,.3,.4,.5,.6,.7,.8,.9), ($w_Q=$
-1,-.9,-.8,-.7,-.6,-.5,-.4,-.3,-.2,-.1,-.01) \& ($\tau_c =$0, .025,
.05, .075, .1, .15, .2, .3, .5) when $\Omega_k$=0. For both cases,
($\omega_c =$ .03, .06, .12, .17, .22, .27, .33, .40, .55, .8),
($\omega_b =$ .003125, .00625, .0125, .0175, .020, .025, .030, .035,
.04, .05, .075, .10, .15, .2), ($n_s = $1.5, 1.45, 1.4, 1.35, 1.3,
1.25, 1.2, 1.175, 1.15, 1.125, 1.1, 1.075, 1.05, 1.025, 1.0, .975,
.95, .925, .9, .875, .85, .825, .8, .775, .75, .725, .7, .65, .6, .55,
.5), $\sigma_8^2$ was continuous, and there were a number of
experimental parameters for calibration and beam uncertainties.
\bibitem{mnu}The simplest interpretation of the
superKamiokande data on atmospheric $\nu_\mu$ is that
$\Omega_{\nu_\tau} \sim 0.001$, about the energy density of stars
in the universe, which implies a cosmologically negligible effect.
Degeneracy between \eg $\nu_\mu$ and $\nu_\tau$ could lead to the
cosmologically very interesting $\Omega_{m\nu}\equiv \Omega_{hdm}
\sim .1$, although the coincidence of closely related energy
densities for baryons, CDM, HDM and dark energy required would be
amazing.
\end{thebibliography}
\end{document}